\documentclass[conference]{IEEEtran}
\IEEEoverridecommandlockouts
\usepackage{cite}
\usepackage{amsmath,amssymb,amsfonts}
\usepackage{algorithmic}
\usepackage{url}
\usepackage{hyperref}
\usepackage{graphicx}
\usepackage{textcomp}
\usepackage{xcolor}
\usepackage[ruled,vlined]{algorithm2e}
\def\BibTeX{{\rm B\kern-.05em{\sc i\kern-.025em b}\kern-.08em
    T\kern-.1667em\lower.7ex\hbox{E}\kern-.125emX}}

\DeclareMathOperator*{\argmax}{argmax}

\begin{document}

\title{Evolution of a Complex Predator-Prey Ecosystem on Large-scale Multi-Agent Deep Reinforcement Learning}

\author{\IEEEauthorblockN{Jun Yamada}
\IEEEauthorblockA{\textit{University College London, Emotech Labs}\\
ucabjy1@ucl.ac.uk}
\and
\IEEEauthorblockN{John Shawe-Taylor}
\IEEEauthorblockA{\textit{University College London}\\
j.shawe-taylor@ucl.ac.uk}
\and
\IEEEauthorblockN{Zafeirios Fountas}
\IEEEauthorblockA{\textit{Emotech Labs, University College London}\\
\href{https://orcid.org/0000-0002-6312-3409}{orcid.org/0000-0002-6312-3409}}
}

\maketitle

\begin{abstract}
Simulation of population dynamics is a central research theme in computational biology, which contributes to understanding the interactions between predators and preys. Conventional mathematical tools of this theme, however, are incapable of accounting for several important attributes of such systems, such as the intelligent and adaptive behavior exhibited by individual agents. This unrealistic setting is often insufficient to simulate properties of population dynamics found in the real-world. In this work, we leverage multi-agent deep reinforcement learning, and we propose a new model of large-scale predator-prey ecosystems. Using different variants of our proposed environment, we show that multi-agent simulations can exhibit key real-world dynamical properties. To obtain this behavior, we firstly define a mating mechanism such that existing agents reproduce new individuals bound by the conditions of the environment. Furthermore, we incorporate a real-time evolutionary algorithm and show that reinforcement learning enhances the evolution of the agents' physical properties such as speed, attack and resilience against attacks.
\end{abstract}


\section{Introduction}
The mathematical simulation of complex dynamical systems \cite{10.2307/j.ctt24hqkz} such as predator-prey ecosystems has been a central problem in understanding intelligent behavior \cite{doi:10.1086/413488} and global dynamics in biology. For instance, the widely-studied Lotoka-Volterra model \cite{10.2307/43430362} is known to produce ordered dynamics of a collection of individual agents in an ecosystem. Such abstract models, typically do not take into account changes in individual behavior and, thus, they are not able to scale well, or to account for animal groups that adapt to new environmental conditions, such as a new terrain. In real ecosystems, there is a continuous competition between the ability of predators to adapt to the new techniques of preys, which try to defend, flee or hide, and vice versa. However, modeling such complex behavior has been until recently very challenging and mathematical models that aim to represent real-world dynamical systems are often over-simplified. In order to understand such complex systems, we need more sophisticated and well-designed simulations where agents employ cognitive abilities. In this work, we present a realistic predator-prey simulation, where agents are trained by reinforcement learning (RL)  \cite{Sutton:1998:IRL:551283} and evolved over multiple generations, resulting in a more biologically approximation of such environments.

Leveraging the power of deep learning, RL has shown outstanding results in various domains, such as Atari video games~\cite{DBLP:journals/corr/MnihKSGAWR13}, the board game Go~\cite{Silver_2016}, robotics~\cite{8675643, DBLP:journals/corr/TaiL16a}, and modeling behavior of biological agents\cite{Dayan2008}. In an attempt to exploit this success in real-world applications, where multiple sources of decision-making are often present, researchers have recently focused on multi-agent reinforcement learning (MARL) systems~\cite{DBLP:journals/corr/abs-1812-11794}.
Such systems are typically applied in environments where different types of agents compete with each other simulating, for instance, a predator-prey ecosystem~\cite{10.2307/1313476, tampuu2015multiagent, DBLP:journals/corr/abs-1710-03748}. However, due to current computational constraints, the number of agents considered in existing research tends to be very limited, seldom exceeding a population of 10 individuals. Hence, most existing systems are unable to capture important dynamical properties observed in real-world populations, a crucial issue that recent work attempts to overcome~\cite{DBLP:journals/corr/abs-1712-00600, DBLP:journals/corr/abs-1709-04511}. Here, we continue this effort, and we use a population of agents on the order of more than 1000 that either compete or co-operate to survive through generations.

In this work, we design the first model where both predator and prey can exhibit unique intelligent behavior using RL in a realistic environment. Specifically, we incorporate an evolutionary strategy \cite{Beyer:2002:ESN:584639.584641} to optimize a number of agent properties over generations and observe how this process is enhanced by the existence of within-lifespan learning. We also integrate a mating mechanism where agents reproduce a new agent under specific conditions within the environment. Finally, we empirically assess the biological realism of the resulting system and we find that key dynamical properties of real predator-prey systems can be replicated.

\section{related work}
\subsection{RL in Predator-prey Ecosystems}

In spite of the recent attention to RL and the clear potential of applying state-of-the-art methods to model animal behavior, the literature so far contains only several studies that employ RL agents to model predator-prey interactions~\cite{schrum:tech08,DBLP:journals/corr/abs-1709-04511, OLSEN2015118,drl_learning_based, tsoularis2015}. 

Yang et al. \cite{DBLP:journals/corr/abs-1709-04511} conducted the first large-scale deep MARL research considering many agents to investigate population dynamics of predator-prey simulated ecosystems. In their work, they showed that the population dynamics of agents exhibit cyclic dynamics similar to the Lotka-Volterra model \cite{10.2307/43430362}, if an RL algorithm controls only predators. Since their preys follow random policy, this oversimplified setting is incapable of modeling fundamental properties of realistic population dynamics. Moreover, their proposed model fails to capture partial observability of the environment because of lack of recurrent neural networks in their proposed model. As underlying dynamics of their environment is a POMDP, deep Q-learning used in their work is insufficient to train agents successfully. Therefore, in our work, we extend deep recurrent Q-learning \cite{DBLP:journals/corr/HausknechtS15} to MARL and train both types of agents to fill this gap. 

\subsection{Population Biology}

Self-organization is a well-studied process in the field of population biology~\cite{ashby1962principles}. It concerns the emergence of globally ordered population dynamics in space and time, realized from collective interactions between agents without any external intervention. In environments where predators and preys interact with each other, the most widely-studied model that captures population dynamics is the one proposed by Lotka and Volterra in \cite{10.2307/43430362}.

This model explains that the population sizes of predators and preys in real environments have a tendency to oscillate with an approximately $90$ degree lag in the phase space. The rate by which these two populations change over time is governed by a non-linear differential equation called \textit{Lotoka-Volterra equation} \cite{10.2307/43430362}: 

Although the equation is able to capture some important population properties, it is based on a number of extreme simplifications of the real world, such as a continuous food supply, genetic stability and limitless appetite, which make them inadequate to account for more complicated dynamical phenomena, such as the existence of quasi-cycles \cite{doi:10.1086/413488}.

Quasi-cycles refer to stochastic oscillatory fluctuations in population size, whose amplitude has the tendency to expand over time due to the existence of a resonance effect. This behavior is expected to be observed whenever a continuous system shows a stable focus. Pineda-Krch et al. \cite{doi:10.1111/j.2006.0030-1299.14940.x} indicated existence of quasi-cycles in a stochastic birth-death process in a predator-prey model. This study also suggested applying the auto-correlation function to distinguish between noisy nodes, quasi-cycles, and noisy limit cycles. If a given population exhibits quasi-cycles, a weak oscillation is presented after applying the auto-correlation function, as opposed to other types of dynamics exhibiting strong oscillations.

Learning in predators and preys has been studied from various perspectives, such as \cite{10.1086/303202, doi:10.1139/z90-092}. The interaction between these two types of animals is considered as sequential events such that preys attempt to escape from predators when they encounter each other, while predators try to capture the former. As a result of the event, the preys face the increasing of predation risk \cite{DefenseAgainstPredators}. Activity of predators is different depending on space and time. This leads preys to require a balance between foraging and reproduction with the risk of predation. That is, preys that show less avoidance response against predators have less mortality than the ones with sensitive avoidance response. Therefore, the preys which have sensitive response against predators are likely to survive longer \cite{10.2307/4600242}.

\section{Methods}

\subsection{Predator-Prey Environment}
\begin{figure}[ht]
  \begin{center}
      \includegraphics[width=8cm]{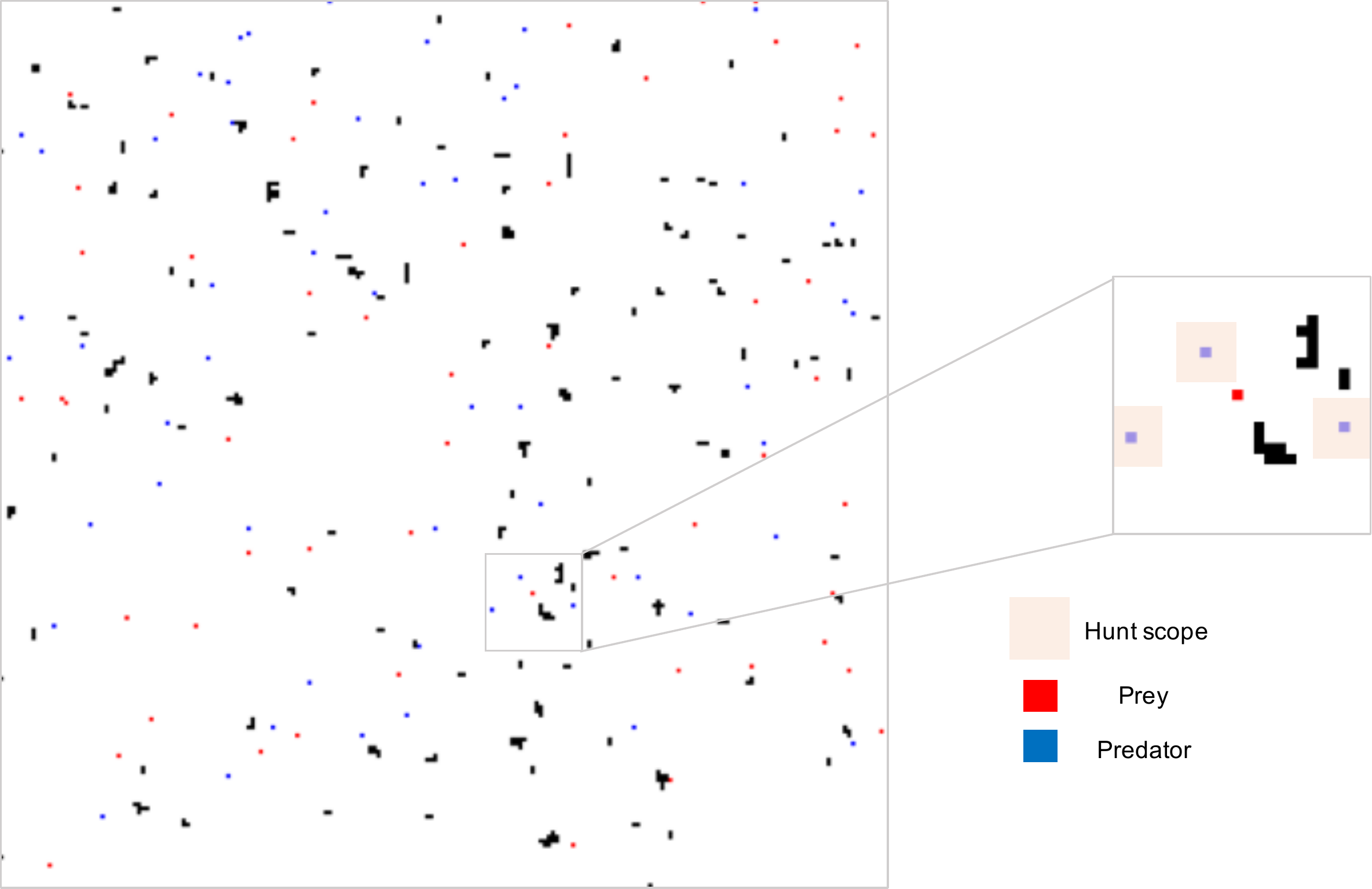}
  \end{center}
  \caption{Environment of a predator-prey ecosystem for experiments (\textbf{Left}). A partial environment with predation square of predators (\textbf{Right}). A red square and a blue square represent a prey and a predator respectively. Black squares are walls. Each predator has predation square where the predator can capture the prey within the square.}
  \label{fig:environment}
\end{figure}

In this work, we propose a new large-scale predator-prey environment for MARL illustrated in Figure \ref{fig:environment}. As opposed to the previous work \cite{DBLP:journals/corr/abs-1709-04511}, we can train both predators and preys by RL in this environment. The environment is unbounded, that is, if an agent moves towards outside of a grid world on an edge of the environment, then it appears at a cell on the other side of edge of the grid, since real-world environments are generally not closed. 

Each predator has a health value and consumes a constant value of health for each movement. It gains a constant value of health when capturing a prey. However, if the health value becomes zero, the predator dies due to hunger. On the other hand, we assume infinite appetite for preys similar to the previous work \cite{DBLP:journals/corr/abs-1709-04511}. In other words, they do not decrease their health for each movement, and they die only because of the predation. An action space in this environment is $\mathcal{A} = \{\text{righ}, \text{left}, \text{forward}, \text{backward}\}$.

Each agent does not perceive the full state of an environment, as Gomez and Mikkulainen \cite{Gomez:1997:IEC:261496.261500} described that a partial observation is common in a predator-prey environment. Therefore, this task is formalized as a POMDP. This represents a special form of non-Markov problem in which the state representation does not have the Markov property, but the underlying dynamics can be considered as MDPs. In this environment, each agent receives the same size of square visual observation. 

Due to the POMDP, each agent receives different observations, depending on its current location in the environment. Objects such as predator, prey, wall and empty cell on a grid world are represented as raw RGB pixels. In addition to a raw visual observation, agents receive other information such as health value. The number of input dimensions is contingent on the type of experiments. However, the first three channels are always occupied by RGB pixel values of objects in a perception scope of agents for all the experiments.

For each timestep, a location of each agent is updated simultaneously. After all agents move to new positions, predators check whether there are any preys in $5\times5$ grid cell of predation scope to hunt the preys. The predators are located at the center of the predation square. If there are any preys in the area, the closest prey from the predator is killed. The predator cannot kill more than one prey at the same timestep. In what follows, we introduce three types of environments we developed for the experiments.

\subsubsection{Environment 1} \label{sec:scenario1_env}
In this environment, both new predators and preys are generated according to a constant ratio, against the total number of either type of agents for each timestep, such that 
\begin{equation}\label{eq:increasing}
N_{t}^{new\_agent} = \min(1, N^{predator (prey)}_{t} \times p^{predator (prey)})    
\end{equation}
where $N_{t}^{new\_agent}$ is the number of new agents , $N^{predator(prey)}_{t}$ is the number of predators or preys at timestep $t$ in the environment and $p^{predator(prey)}$ is the constant ratio to generate new agents. In this environment, it ensures that at least one new agent is added to the environment. Reward functions for agents are formulated as follows:
\begin{equation} \label{eq:reward_predator}
reward_{predator} = \begin{cases}
1, \text{if a predator captures a prey}\\
0, \text{otherwise} \\
\end{cases}
\end{equation}
\begin{equation} \label{eq:reward_prey}
reward_{prey} = \begin{cases}
-1, \text{if a prey is captured by a predator}\\
0, \text{otherwise} \\
\end{cases}
\end{equation}

In this environment, preys only learn how to avoid predation, and predators acquire a hunting skill. Initial health of the predators are randomly sampled from uniform distribution on $[0.5, 1]$. Initial position of agents are randomly chosen from available grid coordinates.

An observation of each agent $i$ consists of 4 channels $o^{i}\in \mathcal{R}^{n \times m \times 4}$, where $n$ is height and $m$ is width of the observation. The last channel represents health of agents, which is crucial information for predators.

\subsubsection{Environment 2}\label{sec:scenario2_env}

In this environment, a new agent is reproduced by process of mating. Two agents mate with a certain probability, defined as a \textit{mating probability}, which is a function of the proximity between agents. An agent needs to be in a certain area of other agents where those agents are located at the center of the square mating scope. The scope for the mating is defined as a constant parameter. Agents cannot mate with multiple agents at the same timestep. If they succeed in mating, then they obtain positive rewards. Reward functions for predators and preys are represented as follows
\begin{equation}
reward_{predator} = \begin{cases}
1, \text{if a predator captures a prey}\\
4, \text{if a predator produces a new predator} \\
0, \text{otherwise} \\
\end{cases}
\end{equation}
\begin{equation}
reward_{prey} = \begin{cases}
-1, \text{If a prey is captured by a predator}\\
4, \text{If a prey reproduces a new prey} \\
0, \text{otherwise} \\
\end{cases}
\end{equation}

Furthermore, we add one agent for predator and prey for each timestep in order to prevent the population of agents from being zero. 
In this environment, both agents need to consider balance between mating and their other habits \cite{DefenseAgainstPredators} such as predation and avoidance. In other words, if a prey focuses on reproduction too much, then its predation risk increases. On the other hand, if it keeps escaping from hostile agents, then the prey cannot obtain higher positive reward from the mating. Therefore, both types of agents need to learn more complex interaction compared to Environment 1. 

A structure of the observation for each agent is the same as the one in Environment 1. In this experiment, the health status is more critical because predators need to balance between predation and mating. If the health of the predator is small, then the predator needs to focus on the predation more than mating; otherwise, the predator dies because of hunger.

\subsubsection{Environment 3}
In Environment 3, we incorporate an evolutionary algorithm into a predator-prey ecosystem to optimize real-value parameters which represent physical abilities of agents. For each predator, \textit{speed} and \textit{attack} are defined as real-value parameters. On the other hand, preys have parameters of \textit{speed} and \textit{resilience} against the attack of predators. In this environment, we do not consider a mating mechanism used in Environment 2.
If a predator attacks a prey, then a resilience value decreases with its value of the attack.
If the resilience becomes zero, then the prey dies. Moreover, if the prey is killed by several predators at the same time, then reward is shared among the several predators. Reward functions are the same as Environment 1.

Similar to Environment 1, the number of new agents is determined according to Equation \ref{eq:increasing}. However, when new agents are generated, a pair of two agents are randomly chosen as parents from all the individuals, and parameters of the two agents are recombined for the new agent, which corresponds to selection and crossover \cite{Reeves:2002:GAP:863042} in an evolutionary algorithm. Since parameters are defined as real values, we use an evolution strategy for the recombination of parameters. Assuming that we have two agents $A$ and $B$ which have a parameter of $p_{A}$ and $p_{B}$ respectively. Then, a parameter of a new agent $C$ is defined as $p_{C} = r p_{A} + (1-r)p_{B}$, where $r$ is a ratio sampled from the uniform distribution on $[0, 1]$. We consider that a fitness function for agents is how long they have survived. This assumption is inspired by the fact that the agents should survive longer if they have better physical properties, and they have more chances for reproduction. Therefore, better set of parameters are more likely to be chosen in long term perspective through evolution of agents. During recombination of parameters, a random value sampled from normal distribution with mean of $0$ and variance of $1$ is added to the recombined value with a certain probability for mutation. This produces versatile sets of parameters for agents.

An observation of each agent $i$ is composed of 7 channels $o^{i}\in \mathcal{R}^{n \times m \times 7}$. The 4th channel represents health status of agents. The other 3 channels correspond to attack, resilience and speed respectively. In order to kill a prey efficiently, a predator needs to find a prey with a small resilience value. Moreover, the prey has to avoid predators with a high attack value; otherwise, the prey is killed by the predator instantly.

\subsection{RL algorithm}
In this work, to resolve an issue of a POMDP in a predator-prey environment, we extend deep recurrent Q-network \cite{DBLP:journals/corr/HausknechtS15} to a MARL domain to train many agents. Deep Q-learning used in the previous work \cite{DBLP:journals/corr/abs-1709-04511} is unsuitable since it cannot store previous contexts of behavior. 
In addition to this, we assign random values sampled from normal distribution represented as \textit{ID} to each agent. This random variable is used as an input for our model to induce diverse behavior.

\begin{figure}[ht]
  \begin{center}
      \includegraphics[width=7cm]{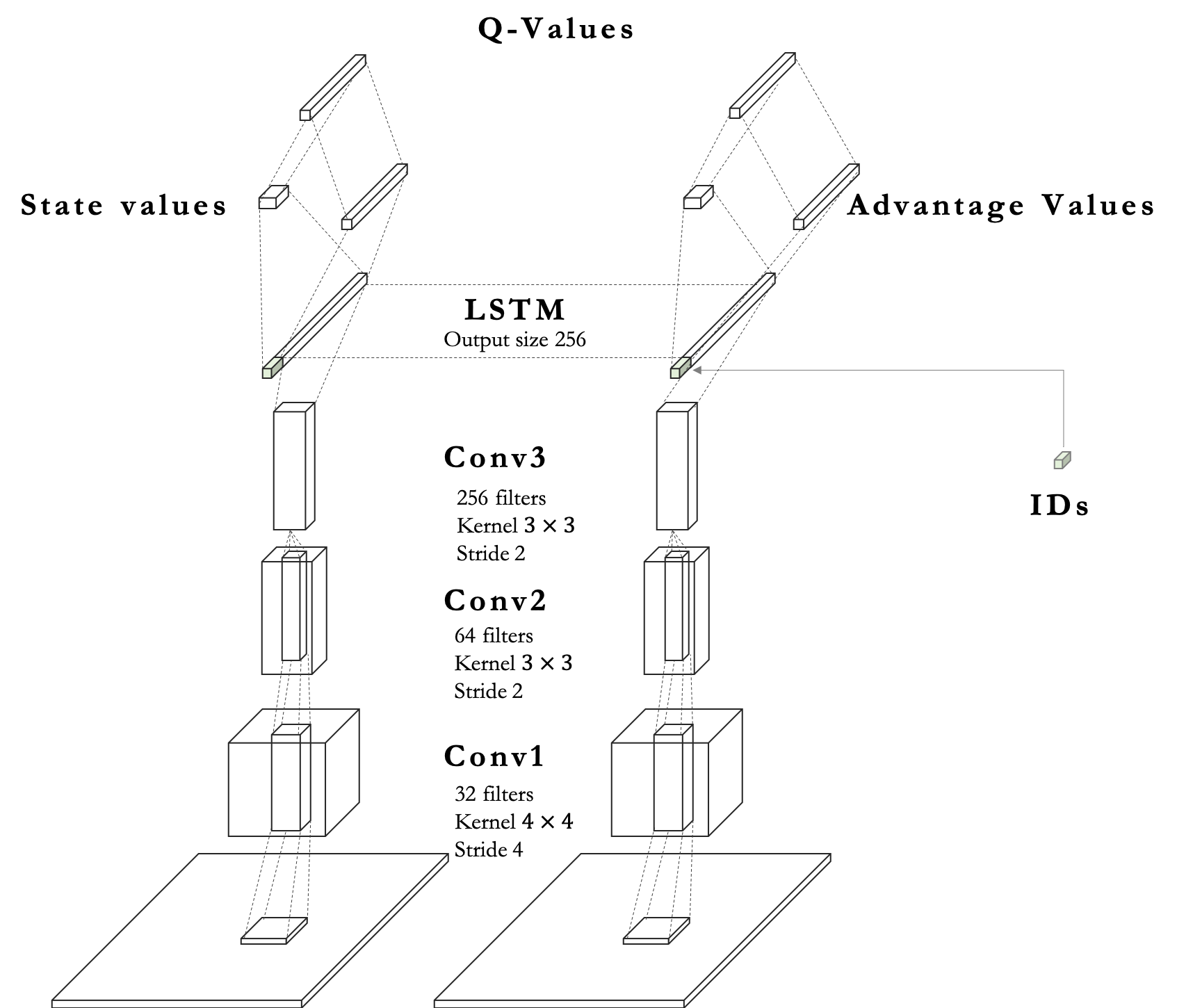}
  \end{center}
  \caption{Architecture of our proposed model. This model is based on Deep Recurrent Q-learning with Dueling neural networks.}
  \label{fig:model_arc}
\end{figure}

As opposed to the prior work\cite{DBLP:journals/corr/abs-1709-04511} using linear layers in their proposed model, we use convolutional neural networks \cite{Lecun98gradient-basedlearning} to capture spatial information for predation, avoidance. \ref{fig:model_arc} is architecture of our proposed model. 
 
\textit{ID} of agents sampled from normal distribution is embedded by a linear layer. Then, the features derived from the LSTM layer are concatenated with the embedded vector of \textit{ID}, which results in distinguishing each agent implicitly and encourage diverse behavior. We utilize dueling neural network architecture \cite{DBLP:journals/corr/WangFL15} for better policy evaluation. Furthermore, we apply double Q-learning \cite{DBLP:journals/corr/HasseltGS15} to stabilize training by reducing high bias caused by an overestimation of Q-learning. 

State-action value for each agent $i \in \{1, 2, \dots N\}$ is updated in the following
\begin{equation}
\begin{split}
    Q^{\pi}&(s_{t}^{i}, a_{t}^{i}) \leftarrow Q^{\pi}(s_{t}^{i}, a_{t}^{i}) + \\ 
    & \alpha [r_{t}^{i} + \gamma Q^{\pi}(s_{t+1}^{i}, \argmax_{a \in \mathcal{A}} Q^{\pi}(s_{t+1}^{i}, a) )- Q^{\pi}(s_{t}^{i}, a_{t}^{i})]
\end{split}
\end{equation}
where $\alpha$ is a step size and $\mathcal{A}$ is the action space.

In order to train our proposed model, we need to run a simulation for several fixed timesteps to accumulate experience and update recurrent neural networks. However, in our environment, if an agent dies during the fixed timesteps, then we cannot apply the back-propagation because the agent does not exist in the environment when we update the parameters of neural networks. To circumvent this, we duplicate the environment as a virtual environment and run the simulation for the fixed timesteps on the copied environment. In the virtual environment, agents are not removed from the environment even though agents die. Although those agents are not eliminated from the environment at that moment, they receive negative rewards. After the fixed timesteps, we update the parameters by back-propagation and discard the virtual environment. 
We continue this process during training. In our experiment, we need to initialize hidden states and cell states for LSTM when a virtual environment is initialized. In the original environment, hidden states and cell states are initialized only when new agents are generated, then the hidden states and cell states are updated and stored until those agents die. If the agents are eliminated from the environment, those states and ID are discarded from memory. During training, we use $\epsilon$-greedy policy as behavior policy.

\section{Experimental Results}
In our experiments, we initialize an environment of size $600 \times 600$. Initial population of each type of agents is $1000$ for training. Square scope for predation is $5\times5$.

\subsection{Verification of agents' intelligence}
\begin{figure}[ht]
    \makebox[0pt]{
    \includegraphics[width=\linewidth]{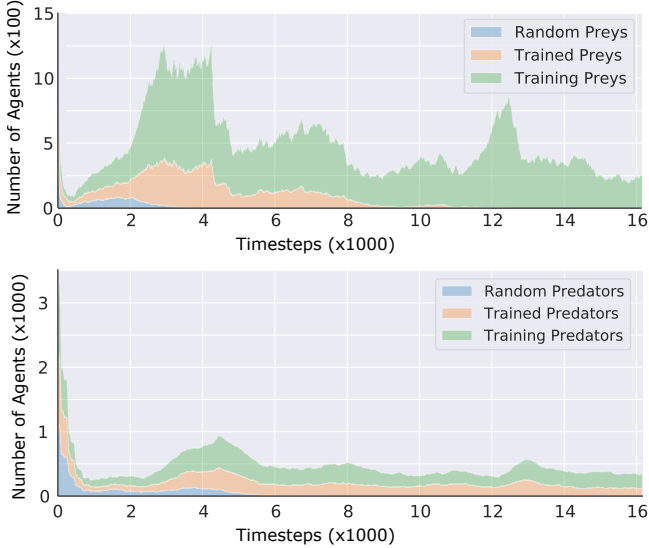}}
    \centering
  \caption{Proportion of the population among three types of agents: agents with random policy, fixed pre-trained weights, and continual learning based on pre-trained weights for preys ({\bf Top}) and predators ({\bf Bottom}).}
  \label{fig:three_type_agents}
\end{figure}

In a large-scale MARL, analysis of whether agents are effectively trained or not is challenging due to the large number of the agents and interactions in an environment. For the same reason, visualizing the interaction between agents are not sufficient to verify it. Therefore, we analyze learning ability by observing how the population change over time among three types of agents: agents with random policy, trained agents with fixed parameters of neural networks, and trained agents with continual learning. Ideally, smarter agents survive and dominate other types of those in an environment. For example, a prey with random policy does not have any cognitive ability to protect itself from attack of hostile agents, leading to extinction. Furthermore, those agents are less likely to reproduce new agents due to the same reason. Figure \ref{fig:three_type_agents} represents the proportion of three types of agents in predators and preys in Environment 2 using a mating mechanism. The initial number of agents is $1000$ for each type. In this experiment, if parents of a new agent do not have the same policy type, for example, one agent has trained policy and the other one is controlled by random policy, then a policy type of the new agent is either trained policy or random policy with the probability of $0.5$. If policy types of parents are the same, then the new agent also has the same policy type of the parents. According to Figure \ref{fig:three_type_agents}, preys following random policy are exterminated. At later timesteps, all of the trained preys with fixed parameters are eliminated, then the one with continual learning dominates all of the other preys. Thus, this implies that we can expect improvement in policy of preys. Also, as the preys with a random policy are eliminated soon, we can assume that the policy is successfully trained to escape from predators and reproduce new agents. On the other hand, although predators with random policy are eliminated soon as well, the population of the other two types of predators remains mostly equal. This implies that the policy of the ``training'' predators has converged.

\subsection{Evolution facilitated by RL}
\begin{figure}[ht]
  \makebox[0pt]{
      \includegraphics[width=6.5cm]{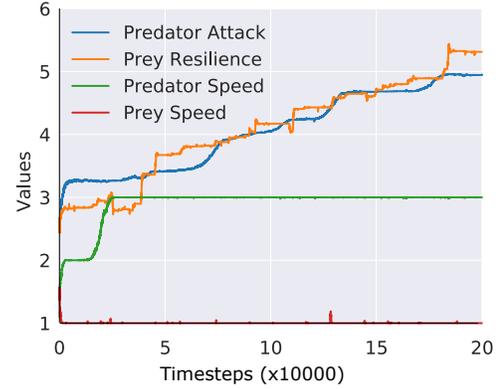}}
  \centering
  \caption{Transition of averaged values of parameters defined in each agent with RL policy during test time}
  \label{fig:genetic_diff}
  \end{figure}

\begin{figure}[ht]
  \makebox[0pt]{
      \includegraphics[width=7.5cm]{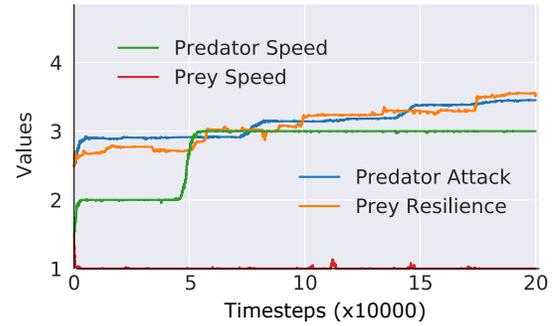}}
  \centering
  \caption{Transition of averaged values of parameters defined in each agent with random policy during test time.}
  \label{fig:genetic_diff_random}
  \end{figure}

We first show the result of an experiment in Environment 3, where an evolutionary algorithm is incorporated into a predator-prey ecosystem to observe agents' physical evolution. The result of this experiment indicates that the average values of parameters of attack and resilience are competitively improved and enhanced by RL. In this experiment, increasing rates for predators and preys $p^{predator(prey)}$ in Equation \ref{eq:increasing} are 0.003 and 0.006 respectively. Figure \ref{fig:genetic_diff} and \ref{fig:genetic_diff_random} illustrate the transition of the average value of each parameter over the population of agents with RL and random policy respectively. According to Figure \ref{fig:genetic_diff}, the average value of attack of predators and resilience of preys keep being improved competitively. After 200000 timesteps, the average attack and resilience achieve around $5$ and $5.5$ respectively to hunt preys or protect themselves from enemies in RL. On the other hand, Figure \ref{fig:genetic_diff_random} shows that the average values of both attack and resilience of agents with random policy take around 3.5 over the same timesteps as policy trained by RL. Moreover, even though those parameters get competitively and gradually improved by random policy, the rate of improvement is slower than the transition with RL. Therefore, these results indicate that behaviors determined by RL enhance genetic evolution of agents. As this assumption is caused by frequent interaction between predators and preys, we could also infer that RL works well in this environment. 

\subsection{Robust Lotka-Volterra dynamics}
\begin{figure}[ht]
      \includegraphics[width=\linewidth]{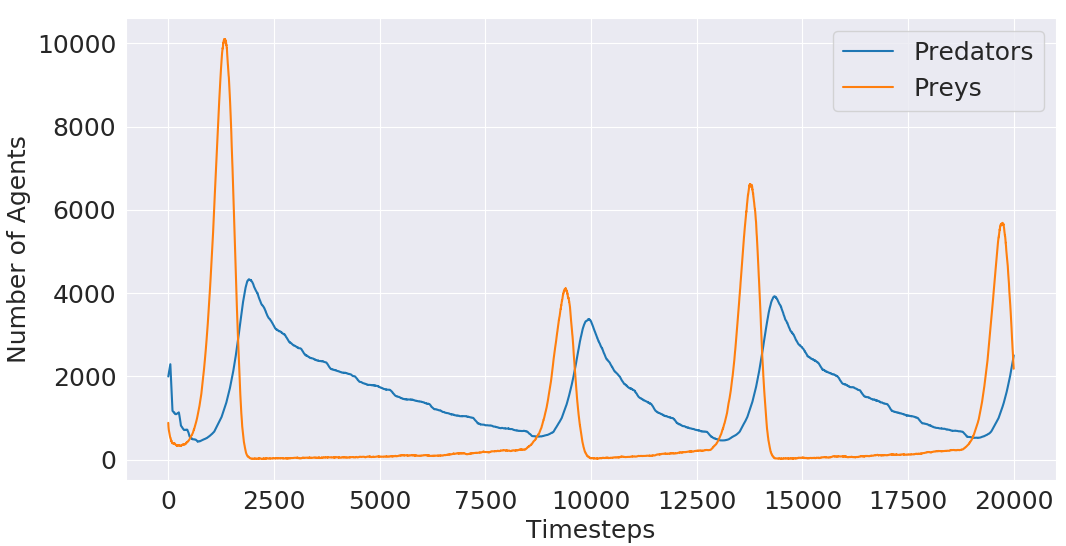}
    \centering
  \caption{Population dynamics of agents with random policy in Environment 1.}
  \label{fig:random_agent}
\end{figure}

The result of this experiment shows that population dynamics in Environment 1 indicates stable cyclic coexistence dynamics similar to Lotka-Volterra model and stable equilibrium coexistence dynamics. In this experiment, we use the same parameters for increasing rate of reproduction for both types of agents as the experiment in Environment 3. Figure \ref{fig:random_agent} shows the population dynamics of agents with random policy. According to Figure \ref{fig:random_agent}, the interaction of agents without intelligence even show unstable cyclic dynamics. However, the cyclic dynamics is disordered as the population does not achieve peaks for fixed timesteps, and the population of those peaks is unsteady. Moreover, the number of preys is often close to zero for a long time before it increases. This assumption is interpreted as a result of population density. In other words, if the environment is mostly filled with preys, then predators can kill the preys easily without any intelligence due to the high density of the preys. Likewise, when the population density of the preys is low, the predators cannot survive because there are not enough preys to maintain their health, which causes an increase in the population of the preys. As a result of these, this demonstrates cyclic dynamics to some extent, however, the population dynamics is still inordinate.

\begin{figure}[ht]
    \makebox[0pt]{
      \includegraphics[width=\linewidth]{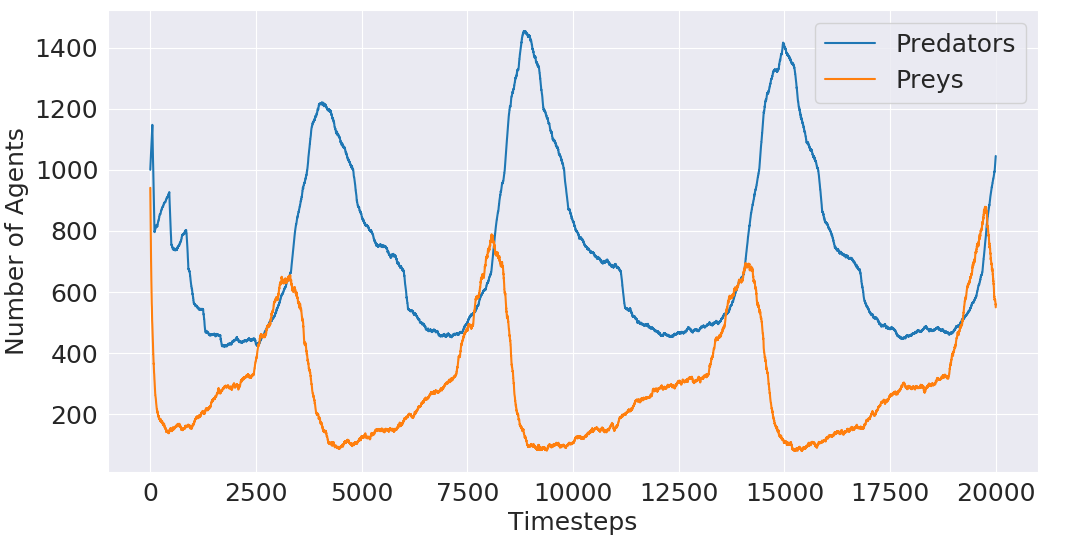}}
    \centering
  \caption{Population dynamics of agents with RL policy in  during test time Environment 1.}
  \label{fig:trained_agent}
\end{figure}

\begin{figure}[ht]
  \begin{center}
      \includegraphics[width=6.5cm]{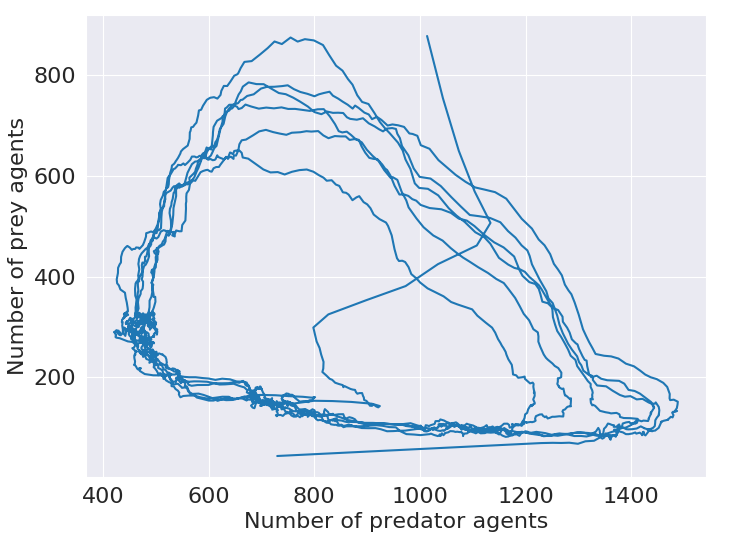}
  \end{center}
  \caption{Relationship between the population of predators and preys in Figure \ref{fig:trained_agent}. As the population dynamics shows the stable cyclic coexistence dynamics similar to the Lotka-Volterra model, each circle is mostly overlapped.}
  \label{fig:circle}
\end{figure}

On the other hand, Figure \ref{fig:trained_agent} represents the population dynamics of agents trained by our proposed model. Therefore, preys try to reduce the predation risk, and predators attempt to kill the preys. According to Figure \ref{fig:trained_agent}, these decision-makings lead to stable cyclic dynamics even before the population density becomes high unlike agents with random policy. The peaks of the population of the predators and preys are $1200$ and $800$ respectively, which are much smaller than the peaks in Figure \ref{fig:random_agent}. Thus, by comparing this with the population dynamics in Figure \ref{fig:random_agent}, we can state that the stable cyclic dynamics is not caused by the high density of the predators or preys if we train agents by RL. Figure \ref{fig:circle} represents the relationship between the population of the predators and preys over time. In Figure \ref{fig:circle}, one circle represents one cycle of the population dynamics in Figure \ref{fig:trained_agent}. As circles are nearly overlapped, the oscillation in Figure \ref{fig:trained_agent} is stable. As a result of this experiment, we could demonstrate that the interaction between trained predators and preys also shows the stable cyclic dynamics similar to Lotka-Volterra model, in comparison to the previous work \cite{DBLP:journals/corr/abs-1709-04511} showing this assumption by training only the predators.

\begin{figure}[ht]
  \begin{center}
      \includegraphics[width=\linewidth]{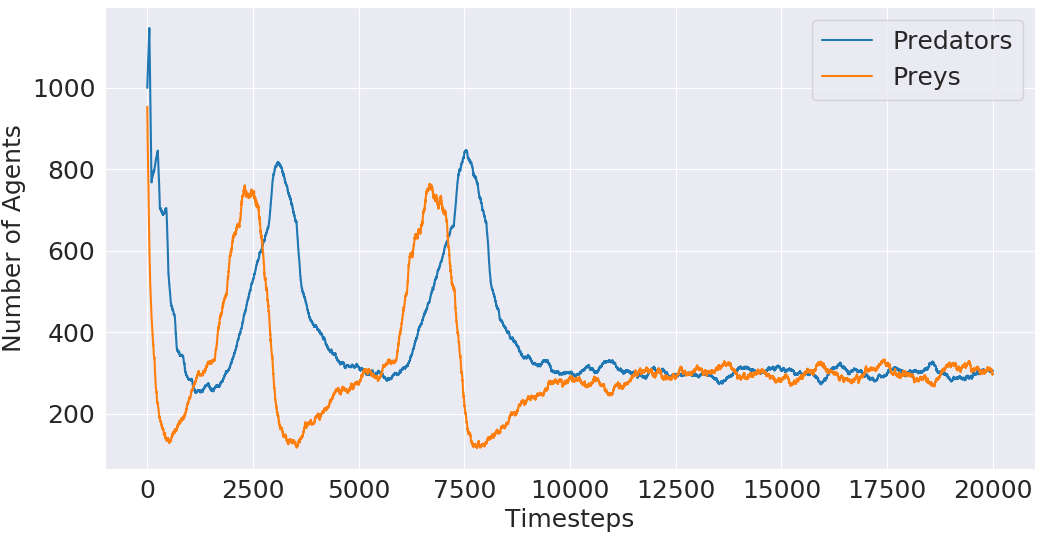}
  \end{center}
  \caption{Stable coexistence dynamics during test time in Environment 1. The trained weight is different from the one used in Figure \ref{fig:trained_agent}. After 10000 timesteps, the population dynamics becomes stable and does not show the cyclic dynamics similar to the Lotka-Volterra Model.}
  \label{fig:stable_coexistence}
\end{figure}

Moreover, by using different trained weights, the population dynamics demonstrates stable coexistence dynamics shown in Figure \ref{fig:stable_coexistence}. According to Figure \ref{fig:stable_coexistence}, during the first 10000 timesteps, it shows the stable cyclic dynamics. However, after then, the population dynamics becomes stable, which indicates the stable equilibrium coexistence, mentioned in \cite{10.2307/1313476} and well mathematically analyzed in \cite{fleischer_pick_your_trade_offs}. This is one of the types of population dynamics for coexistence, when both predator and prey evolve for learning behavior. Since we trained both predators and preys, we think of this assumption as a result of balanced birth-death process. 
Since MARL is hardly able to find the global optima of policy, we cannot state which trained weight of deep neural networks is better than others. However, throughout these results, we can claim that the population dynamics of learned predator and prey shows cyclic dynamics similar to the Lotka-Volterra model or stable coexistence dynamics, which is also verified by a theory of biology \cite{10.2307/1313476}. Also, those results imply that prey successfully learns to escape from predator and predator learn to attack prey.

\subsection{Stochastic birth-death dynamics}
This experiment demonstrates that the population dynamics of trained agents in Environment 2 is similar to quasi-cycles \cite{doi:10.1086/413488} which is the assumption of a predator-prey ecosystem conditioned on stochastic birth-death process. On the other hand, the population dynamics of agents with random policy does not show any cyclic dynamics. In this experiment, mating probabilities for predator and prey are $0.003$ and $0.006$ respectively, and mating square scope is $15 \times 15$ grid cell for both types of agents.

\begin{figure}[h!]
    \makebox[0pt]{
      \includegraphics[width=\linewidth]{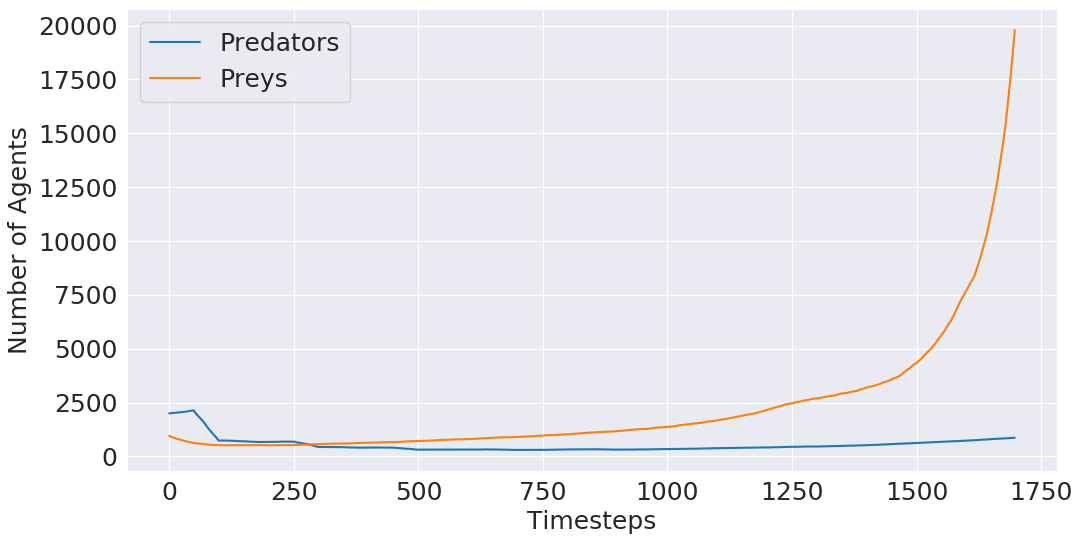}}
    \centering
  \caption{Population Dynamics of random agents in Environment 2. The population dynamics is inordinate since the population of prey keep increasing and does show any stable dynamics.}   
  \label{fig:ga_random}
\end{figure}

Figure \ref{fig:ga_random} is the population dynamics of agents with random policy in Environment 2. Since they do not know how to mate with other agents, the population of preys, which has higher mating probability, rapidly increases due to higher density of the population and the higher mating probability. On the contrary, mating for predators rarely happens when the population of the predator once become small due to small population density, which leads that the predators rarely get close to each other. As a result of this, the population dynamics of the agents taking random action does not show any cyclic oscillation.

\begin{figure}[ht]
  \makebox[0pt]{
      \includegraphics[width=\linewidth]{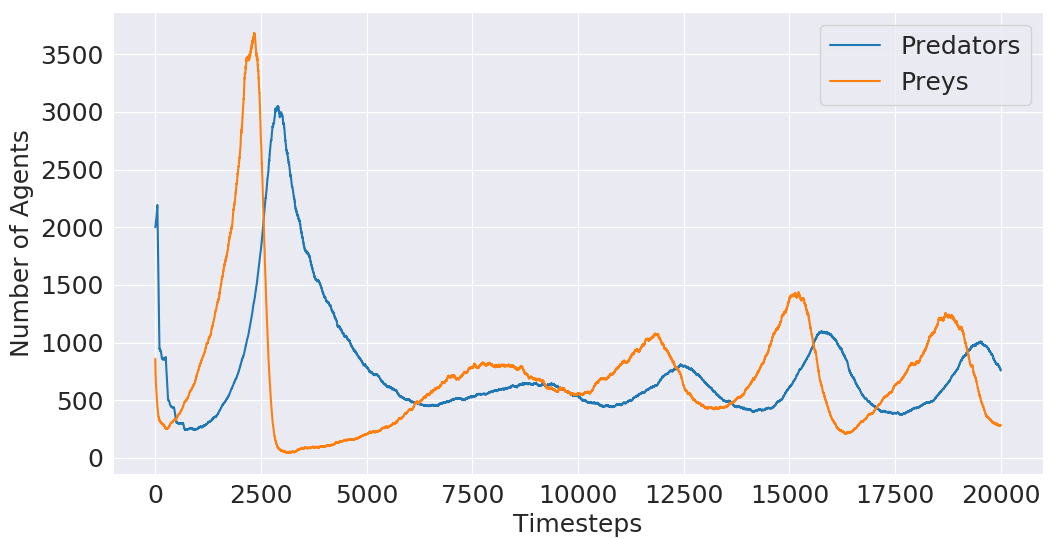}}
  \centering
  \caption{Population Dynamics with the RL during test time in Environment 2. The population dynamics show the cyclic dynamics different from the dynamics of agents with a random policy.}
  \label{fig:ga_plot}
\end{figure}

\begin{figure}[ht]
    \makebox[0pt]{
      \includegraphics[width=6.5cm]{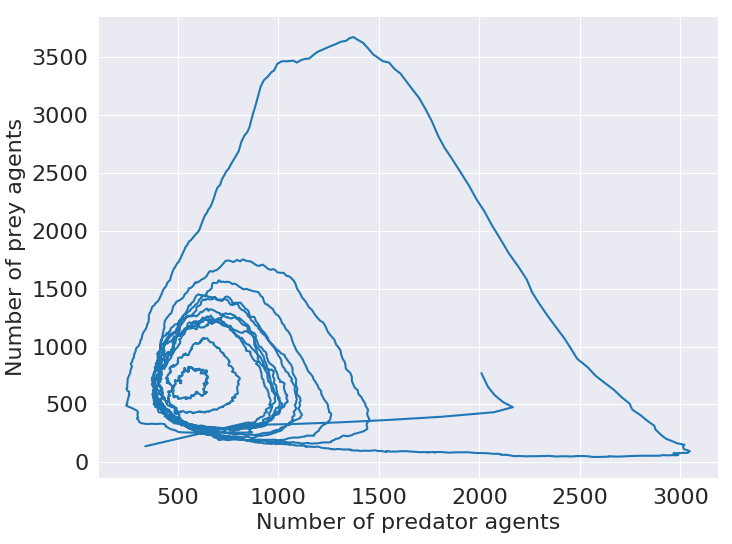}}    
    \centering
  \caption{Relationship of the population between predators and preys during test time in Environment 2. The circles get smaller over time.}
  \label{fig:circle_ga}
\end{figure}

On the other hand, Figure \ref{fig:ga_plot} shows the population dynamics of agents trained with RL. Compared to Figure \ref{fig:ga_random}, Figure \ref{fig:ga_plot} shows cyclic dynamics, as the peaks of the population of predators and preys are interchanged. Therefore, this result implies that the interaction between intelligent agents shows the cyclic dynamics similar to the Lotka-Volterra equation in a more complex environment such that a mating mechanism is incorporated. Furthermore, Figure \ref{fig:circle_ga} illustrates the relationship between the population of the predator and the prey over time. Compared to Figure \ref{fig:circle}, Figure \ref{fig:circle_ga} demonstrates that the circles become smaller as time goes by. To analyze this in detail, we consider analysis of quasi-cycles caused by stochastic birth-death process in the following.

\begin{figure}[ht]
  \makebox[0pt]{
      \includegraphics[width=\linewidth]{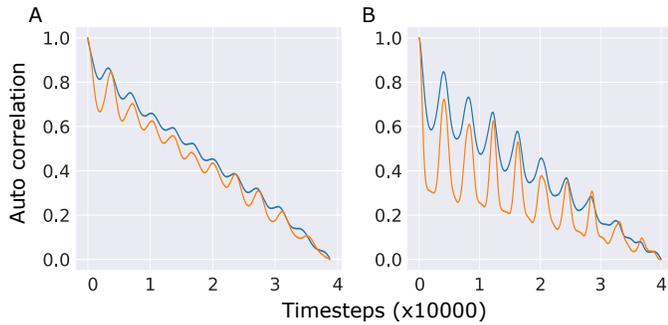}}
  \centering
  \caption{\textbf{(A)} Autocorrelation applied for population dynamics in Environment 2. \textbf{(B)} The same for Environment 1.}
  \label{fig:auto_correlation_both}
\end{figure}

In Environment 2, since a new agent is generated by a mating function and agents die as a consequence of decision-making by RL, birth-death process is considered to be stochastic. In such a case, \cite{75c6bad5f81d4028879fb44272ce7ecd} described that the two populations sizes would follow quasi-cycles. We can distinguish whether the population dynamics exhibit this stochastic pattern or not by applying \textit{autocorrelation function}. Figure \ref{fig:auto_correlation_both}.A shows the result after the autocorrelation function is applied to the population dynamics in Figure \ref{fig:ga_plot}. 
The small oscillation in Figure \ref{fig:auto_correlation_both}.A indicates that the population dynamics is similar to the quasi-cycles. On the other hand, Figure \ref{fig:auto_correlation_both}.B illustrates the result after the autocorrelation function is applied to the result in Environment 1. This autocorrelation is more similar to the behaviour of a stable limit cycle.
Therefore, we can state that the stochastic birth-death process derived from intelligent agents trained by RL exhibit quasi-cycles, as it is shown in biological studies. Furthermore, Figure \ref{fig:circle_ga} indicates a stable focus because the attractor gets towards a certain stable point. That is, the circles which represent the relationship of the population between predators and preys gets smaller as time goes by. This assumption also verifies the quasi-cycles in the population dynamics in Environment 2.

\section{discussion}
In this work, we model key population dynamics normally found in ecosystems using our proposed multi-agent environment. Additionally, we show that evolution of the physical properties used to describe simulated agents in our environment can be enhanced by RL. 

By using or extending our environments, we would compare the population dynamics of agents trained by RL with the ones in theoretical biology. All in all, our proposed model can facilitate future research allowing complex simulations of large-scale predator-prey ecosystems and thus providing unique insights into current questions in biology.

\section*{Acknowledgements}
The authors would like to thank Dr David Murrell for the valuable feedback on the modelling work presented here, as well as Indeera Munasinghe and James Grant for playing a key part in the initial inspiration for this project.

\bibliographystyle{IEEEtranS}
\bibliography{main}

\onecolumn
\appendix
\subsection{Hyperparameters}

\begin{table}[ht]
\begin{tabular}{|l|l|l|l|}
\hline
 & Experiment 1 & Experiment 2 & Experiment 3 \\ \hline \hline
 Environment size (grid cells)& $600 \times 600$ & $600 \times 600$  &  $600 \times 600$ \\ \hline
 discount rate& 0.99 & 0.99 & 0.99  \\ \hline
 initial population (predator) for training & 1000 & 1000 & 1000  \\ \hline
 initial population (prey) for training & 1000 & 1000 & 1000  \\ \hline
 maximum capacity of population for training & 10000 & 10000 & 10000  \\ \hline
 increasing ratio (predator) & $3\cdot 10^{-3}$ & $3\cdot 10^{-3}$ & $3\cdot 10^{-3}$ \\ \hline
 increasing ratio (prey) &$6 \cdot 10^{-3}$ & $6\cdot 10^{-3}$ & $6\cdot 10^{-3}$  \\ \hline
 predation scope (grid cells) & $5\times 5$ & $5\times 5$ & $5\times 5$  \\ \hline
 mating scope (grid cells) &  &  & $15\times 15$  \\ \hline
 mutation probability & $1\cdot 10^{-3}$ &  & \\ \hline
 optimizer & RMSProp & RMSProp & RMSProp \\ \hline
 learning rate & $1\cdot 10^{-4}$ & $1\cdot 10^{-4}$ & $1\cdot 10^{-4}$  \\ \hline
 
\end{tabular}
\end{table}

\subsection{Training algorithm}
\begin{algorithm}[]
\SetAlgoLined
 Initialize agent's Q-network $\pi_{i}$, agent's identity $v_{i}$ \\
 Randomly initialize the environment $s \sim p(\mathcal{S})$ \\
 \For{time step=1,2,\ldots }{
    Initialize hidden states and cell states for LSTM for each agent \\
    copy environment: $s'$ \\
    
    \For{time window $u = 1, 2, \ldots$}{    
        \For {agent $i=1, 2, \ldots$}{
            receive the local observation features $O(i)$ \\
            sample the identity embedding $I(i$) from normal distribution\\
            Take action $a_{t+u}^{i} \sim \pi_{\theta}^{i}(a|s'_{t+u})$ where $\pi_{\theta}^{i}(a|s_{t+u}') = \epsilon\text{-greedy}(Q(s_{t+u}^{'i}, a_{t+u}^{i}))$ \\
            Update copied environment given action $a_{t+u}$ \\
        }
        \If{$u$ is the last time step in the time window}{
            \While {$|\mathcal{B}| \geq$ batch size }{
                Sample a mini batch from $\mathcal{B}$ \\
                Update the parameters of deep neural networks w.r.t. the loss: \\
                    $\delta = (r_{t+u}^{i} + \gamma \max_{a'} Q(s_{t+u+1}^{i}, a') - Q(s_{t+u}^{'i}, a_{t+u}^{i}))^2$
            }
        }
    }
    
    Clear experience replay buffer $\mathcal{B}$ \\
    \For {agent i=1, 2, \ldots}{
        Take action $a_{t}^{i} \sim \pi_{\theta}^{i}(a|s_{t})$ where $\pi_{\theta}^{i}(a|s_{t}) = \epsilon\text{-greedy}(Q(s_{t}^i, a_{t}^{i}))$ \\
        Update the original environment given action $a_{t}^{i}$
    }
    Decrease the health of predators \\
    Assign positive reward to predators who captures prey \\
    Assign negative reward to preys killed by predator \\
    Increase the health of predators if they kill prey \\
    Remove dead agents from the environment \\
    Add new predators and preys to the environments \\
 }
 \caption{Training process in our experiments}
 \label{alg:learning}
\end{algorithm}
\end{document}